\documentclass[aps,a4paper,prb,amsmath,twocolumn,amssymb,titlepage]{revtex4-1}
\usepackage{here,wrapfig, amsmath, amssymb, braket, mathrsfs}
\usepackage{csquotes, amsfonts, multirow, tabularx, array, bm, txfonts, bibentry, ulem, color}
\usepackage{xcolor, etoolbox}
\usepackage{graphicx}

\usepackage{xparse, soul, xcolor}

\NewDocumentCommand{\sotwo}{O{red}O{black}+m}
    {%
        \begingroup
        \setulcolor{#1}%
        \setul{-.5ex}{.4pt}%
        \def\SOUL@uleverysyllable{%
            \rlap{%
                \color{#2}\the\SOUL@syllable
                \SOUL@setkern\SOUL@charkern}%
            \SOUL@ulunderline{%
                \phantom{\the\SOUL@syllable}}%
        }%
        \ul{#3}%
        \endgroup
    }

\newcommand{\kl}{\left(}
\newcommand{\kr}{\right)}

\newcommand{\Loss}{\mathcal{L}}

\newcommand{\vk}{{\bf k}}
\newcommand{\ua}{\uparrow}
\newcommand{\da}{\downarrow}

\newcommand{\ztwo}{\mathbb{Z}_2}

\begin{document} 

\title{Learning Disordered Topological Phases by Statistical Recovery of Symmetry} 
\author{Nobuyuki Yoshioka, Yutaka Akagi, and Hosho Katsura}
\affiliation{Department of Physics, University of Tokyo, 7-3-1 Hongo, Bunkyo-ku, Tokyo 113-0033, Japan}

\begin{abstract}
We apply the artificial neural network in a supervised manner to map out the quantum phase diagram of disordered topological superconductor in class DIII. 
Given the disorder that keeps the discrete symmetries of the ensemble as a whole, translational symmetry which is broken in the quasiparticle distribution individually is recovered statistically by taking an ensemble average. By using this, we  classify the phases by the artificial neural network that learned the  quasiparticle distribution in the clean limit and show that the result is totally consistent  with the calculation by the transfer matrix method or noncommutative geometry approach. If all three phases, namely the $\ztwo$, trivial, and the thermal metal phases appear in the clean limit, the machine can classify them with high confidence over the entire phase diagram. If only the former two phases are present, we find that the machine remains confused in the certain region, leading us to conclude the detection of the unknown phase which is eventually identified as the thermal metal phase. 

\end{abstract}
\maketitle

\section{Introduction}
Machine learning techniques construct and execute the computational algorithm which optimizes the quantified objective defined by gathered training data to make valuable predictions about previously unseen data. The surging development of new techniques has led to the recognition of its effectiveness in various research fields such as condensed matter physics. Examples of previous studies include  the application of the restricted Boltzmann machine to compressed expression of quantum many-body states,~\cite{carleo_2017, torlai_2016,torlai_2017,deng_2016, gao_2017,nomura_2017, carleo_2018, freitas_2018, saito_2017, saito_2018} acceleration of Monte Carlo simulation,~\cite{liu_2017, huang_2017,wang_2017, shen_2018} detection of phase transition by unsupervised learning without teaching the notion of phases to the machine. ~\cite{leiwang_2016, evert_2017, yehua_2017, broecker_unsupervised_2017, hu_2017}
Among them, a problem that draws particular attention is the classification of various phases such as in topological systems,~\cite{carrasquilla_2017, frank_2017, ohtsuki_2016_1, ohtsuki_2017, zhang_shen_2018, carvalho_2018} strongly correlated systems,~\cite{chng_2017, broecker_sign_2017} and many-body localized systems.~\cite{shindler_2017, venderley_2017, evert_2017_2}

In this paper, we investigate the quantum phase diagram of 2d noncentrosymmetric superconductor in class DIII with disorder motivated by the recent proposal of the candidate materials  such as Cu$_x$Bi$_2$Se$_3$~\cite{wray_2010, fu_2010} and FeTe$_x$Se$_{1-x}$.~\cite{wang_2015, wang_2016, zhang_yaji_2018}  The gapless excitation of topological superconductors including class DIII can be described by Majorana edge modes, which attract keen interest from the viewpoint of topological quantum computation.~\cite{kitaev_2003,bravyi_2002} While topological invariants in translational invariant systems have been well studied including their concrete expressions and calculations,~\cite{tknn_1982, kohmoto_1985, kane_mele_2005_1, kane_mele_2005_2} the theoretical understanding in disordered systems is far from complete. In particular, the formulation of Niu-Thouless-Wu, which is an extension to many-body systems and disordered systems,~\cite{niu_1985} is known to  break down for class DIII.

Our goal is to determine the phase diagram for finite disorder by applying the artificial neural network (ANN), given the information of phases only in the clean limit. There are two underlying key concepts. 
The first is the expressibility of the ANN.  While the choice of the appropriate network architecture is a training-data-dependent problem,~\cite{MLTSC_comment6} it is shown that, for arbitrary data groups $\{\vec{x}_i, {\bf F}(\vec{x}_i)\}$ and arbitrary precision $\epsilon>0$, an ANN can be constructed so that its prediction ${\bf \widetilde{F}}$ satisfies $|{\bf \widetilde{F}}(\vec{x}_i) - {\bf F}(\vec{x}_i)|<\epsilon$.~\cite{cybenko_1989, hornik_1989, nielsen_2015}
The second is the  recovery  of the translational symmetry by ensemble average. While the translational symmetry is broken in a system with disorder such as a random potential,~\cite{anderson_1958} as an ensemble of disorder average the symmetry is {\it statistically recovered}.  
Our expectation is that an ANN learned from the data in the clean limit is capable of classifying such ensemble averaged states. 
As we show later, the phase diagram obtained from our method is fully consistent with the results in both the transfer matrix (TM) method~\cite{mackinnon_1983} and the calculation of a $\ztwo$ index by noncommutative geometry which was recently proposed.~\cite{katsura_2016_1, katsura_2016_2, akagi_2017,MLTSC_comment3}

The rest of the paper is organized as follows. In Sec. \ref{methods} we provide the method to map out the phase diagram. The Hamiltonian for 2d noncentrosymmetric  superconductor in class DIII with and without the disorder is introduced here. We also discuss the architecture, input, and output of the ANN which is employed as the classifier. In Sec. \ref{results} we show the results obtained by performing both ternary and binary classification with the ANN, comparing to those by other two methods. Finally, the summary for the current work and the discussion on the future direction  is  given in Sec. \ref{discussions}.  For completeness, we describe the two other methods to depict the phase diagram in Appendices A and B. Also we compare the result with and without the statistical  recovery of symmetry in Appendix C.


%
\section{Methods}\label{methods}
\subsection{Bogoliubov-de Gennes Hamiltonian in real space}
The Bogoliubov-de Gennes Hamiltonian for 2d noncentrosymmetric superconductor in class DIII in the clean limit is given in the real space as~\cite{sato_2009}
\begin{eqnarray}\label{bdgham}
H_0 &=& \sum_{\bf r} \sum_{k=1,2} \Psi^{\dagger}_{\bf r} t_k \Psi_{\bf r+e_k}
+ \sum_{\bf r} \Psi^{\dagger}_{\bf r}v\Psi_{\bf r},\\
t_1 &=& t s_3 \otimes \sigma_0 + \frac{i \Delta}{2}s_1 \otimes \sigma_3,\\
t_2 &=& t s_3 \otimes \sigma_0 + \frac{\Delta}{2} s_1 \otimes \sigma_3,\\
v &=& -\mu s_3\otimes \sigma_0 -\Delta_2 s_2\otimes\sigma_2.
\end{eqnarray}
For concreteness,  our model is defined on a square lattice with cylindrical boundary conditions. Here, $\Psi_{\bf r} = [c_{{\bf r}\ua}, c_{{\bf r}\da}, c^{\dagger}_{{\bf r}\ua}, c^{\dagger}_{{\bf r}\da}]^T$ is the Nambu operator with $c_{{\bf r}\alpha}$ denoting an annihilation operator of an electron with spin $\alpha$ at site $\bf r$, $t_{1(2)}$ and ${\bf e}_{1(2)}$ are the hopping matrix and the primitive vector along the $x(y)-$direction with the transfer integral $t$ and the helical $p$-wave coupling $\Delta$.  The Pauli matrices $s_k$ and $\sigma_k$ ($k$ = 0,1,2,3) operate on particle-hole and spin space, respectively. The on-site term, $v$, consists of the chemical potential $\mu$ and the $s$-wave pairing $\Delta_2$. The mixture of the spin-singlet and the spin-triplet pairings are caused by the broken inversion symmetry.~\cite{MLTSC_comment1}
Note that, in the  Hamiltonian the following symmetries are present: even particle-hole symmetry (PHS), odd time-reversal symmetry, and chiral symmetry.
Thus the topological property is characterized by the $\ztwo$ topological invariant.~\cite{kane_mele_2005_1,kane_mele_2005_2, fu_kane_2007,schnyder_2008, kitaev_2009, qi_2010} For Eq. (\ref{bdgham}),  we find that the system is in the $\ztwo$ phase at $2-2\sqrt{1-(\Delta_2/\Delta)^2} < |\mu| < 2 + 2\sqrt{1-(\Delta_2/\Delta)^2}$ if $|\Delta_2/\Delta|<1$.\cite{diez_2014}
As the on-site disorder, we introduce a random potential with the amplitude distributed uniformly within the width $W$, namely,
\begin{flalign}
& H_W= \sum_{\bf r} \Psi_{\bf r}^{\dagger} (W_{\bf r} s_3 \otimes \sigma_0)\Psi_{\bf r} &\text{for } W_{\bf r} \in [-W/2, W/2].
\end{flalign}
Consequently, the total Hamiltonian takes the form $H = H_0 + H_W$.

Note that once the disorder is turned on, the wave number is no longer a good quantum number and thus the formula for the Kane-Mele invariant is no longer applicable. 
It is known that moderate randomness in spin-rotational symmetry broken system may cause destructive interference of time-reversal paths of the quasiparticle, suppressing the back scattering and thereby leading the system to show metallic behavior (weak-antilocalization) in 2d.~\cite{abrahams_1979, hikami_1981, evers_2008} In particular, ``insulator-metal'' transition from the $\ztwo$ phase, in which Majorana fermions pinned to the disorder percolates, gives rise to the so-called Majorana metal phase.~\cite{senthil_2000} In 2d, the thermal conductivity grows logarithmically with the system size, which is understood as a consequence of the extended behavior of the quasiparticle over the whole system. 
Actually, the metallic property of thermal transport arises also when the bulk gap is closed in the clean limit. 
Thus, all of these will be collectively referred to as the thermal metal (ThM) phase in the following.

\begin{figure}[b]
\begin{center}
\includegraphics[width=0.7\hsize]{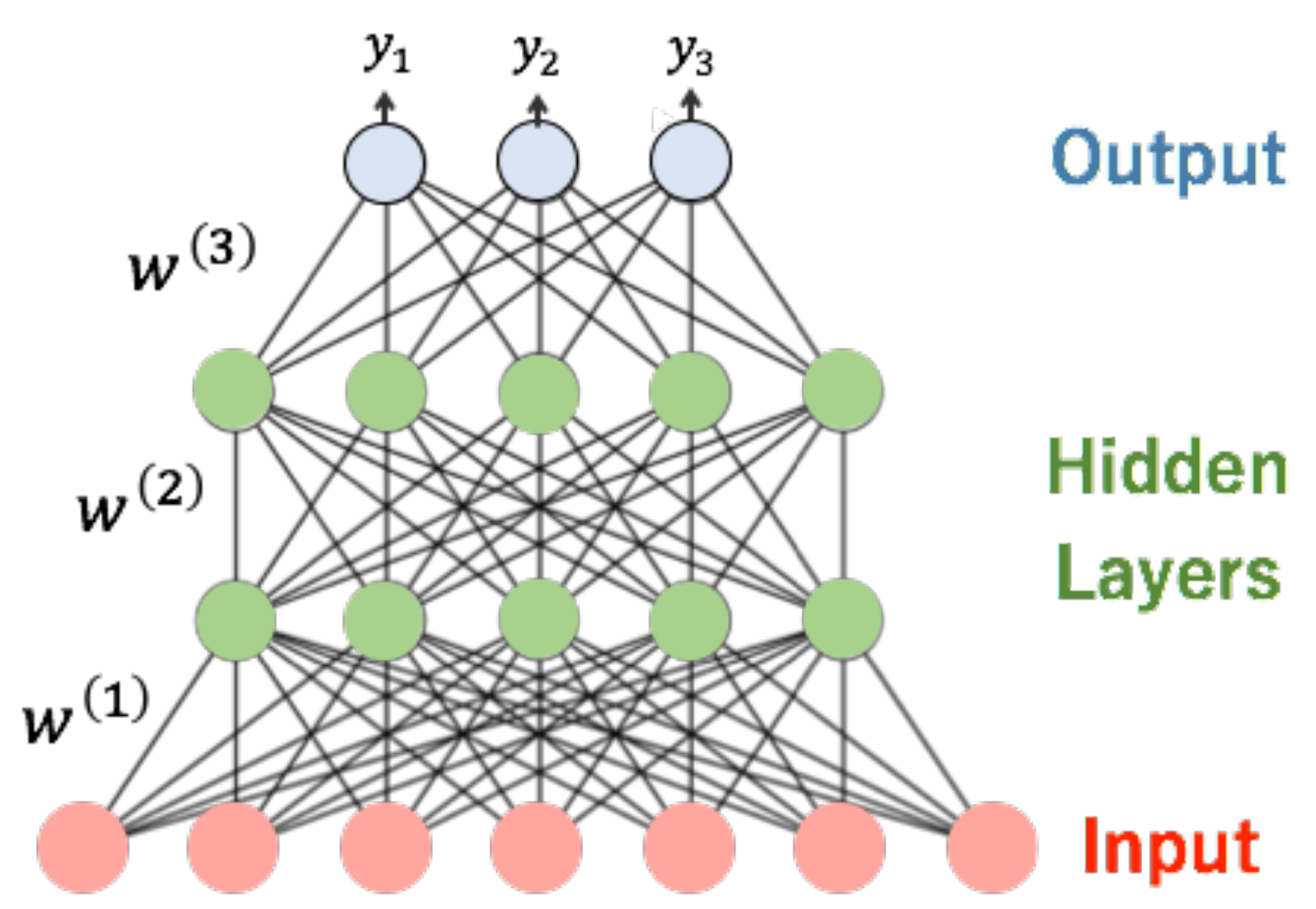}
\caption{(Color online) The architecture of a feedforward artificial neural network  with  two hidden layers, at which the input data is compressed to extract some abstract feature for classification. The activation of the output layer is the Softmax function so that the sum is unity, allowing us to interpret as the confidence of the machine.}  
\label{NN_architecture}
\end{center}
\end{figure}
\begin{figure}[t]
\begin{center}
\includegraphics[width=\hsize]{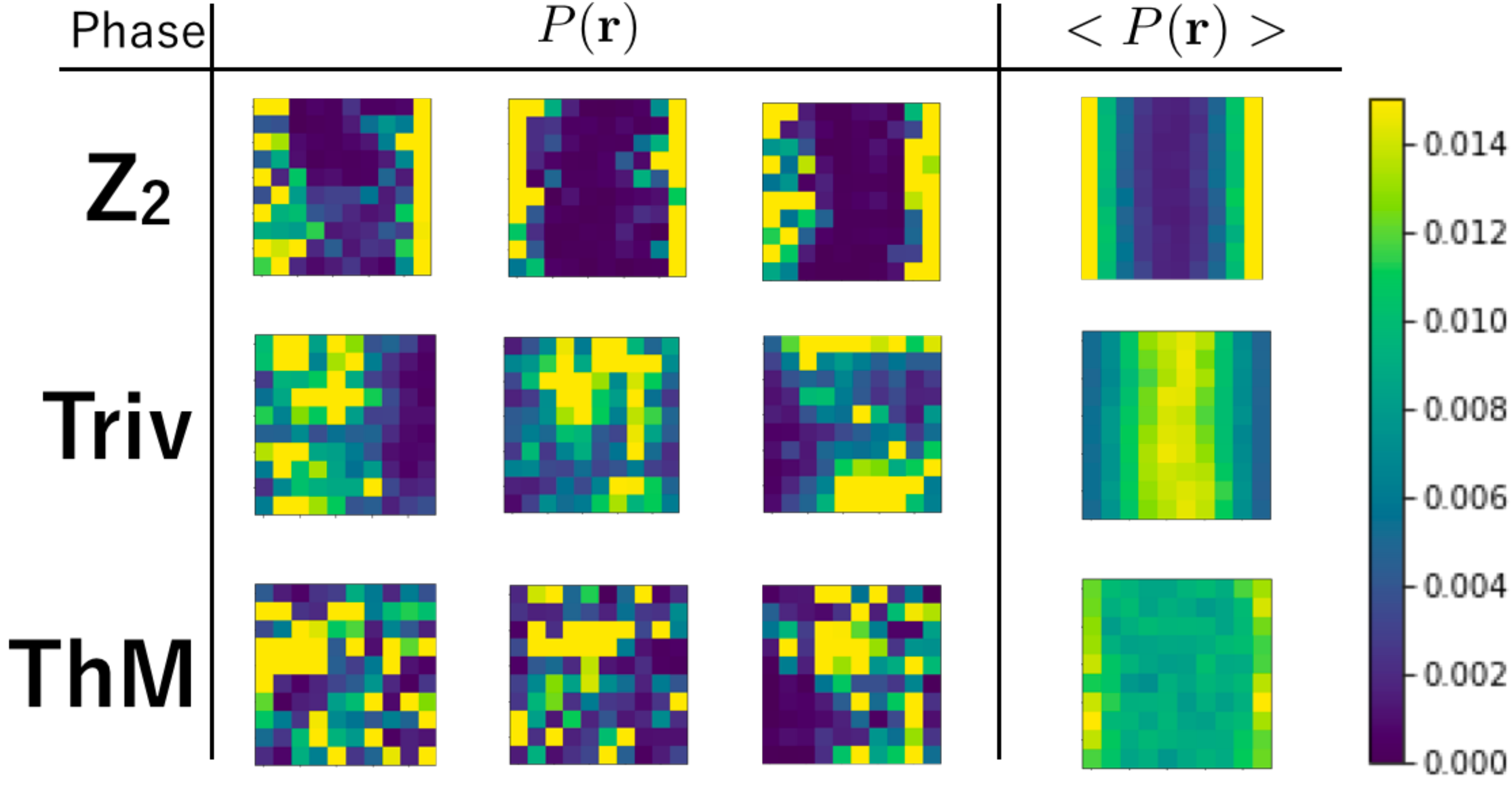}

\caption{(Color online)  Typical single-shot quasiparticle distribution of the first excited state, $P({\bf r})$, and its disorder average, $\braket{P({\bf r})}$,  over 500 realizations of random configurations. The parameters are taken from deep inside the phases as $(\mu,W) = (2,9), (6,5), (2,18)$ with $\Delta = 3$ and $\Delta_2 = 0$ from the top. The system size is taken as $10 \times 10$. }  
\label{dist_av_ex}
\end{center}
\end{figure}
\begin{figure*}[t]
\begin{center}
\includegraphics[width=1.95\columnwidth]{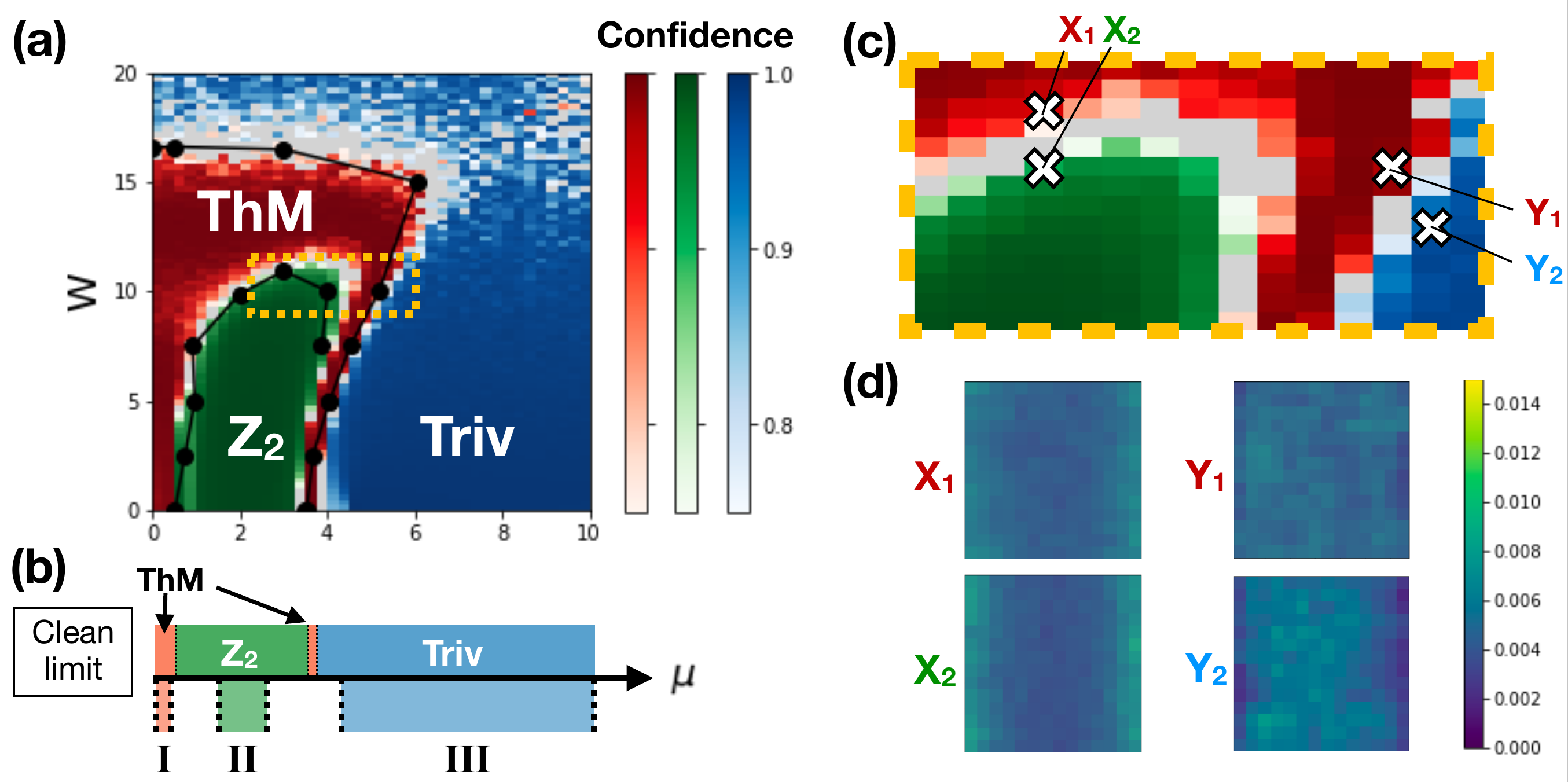}
\caption{(Color online) (a) Average outputs of 200 ternary-classifying ANNs trained with the clean-limit data for $t=1, \Delta = 3, \Delta_2 = 2$. The color of each point $(\mu, W)$ indicates the confidence for the thermal metal (red), $\ztwo$ (green), and  trivial (blue) phase. The machine is highly confident of each phase but confused at the boundary.
(b) The parameter $\mu$ of 1000 training data  with system size $14\times 14$ is uniformly distributed within  I: $[0.0,0.3]$, II: $[1.0,2.5]$, and III: $[4.0,10.0]$.  During the training scheme, the network is tested by the data generated along $\mu \in [0.0,10.0]$ in the clean limit, resulting in accuracy over 90\%. 
(c) Enlargement of the region surrounded by the orange dotted line in (a). (d) The averaged inputs $\braket{P(\bf r)}$ for $N_r = 500$ in the vicinity of the boundaries. The parameters $(\mu, W)$ are given as $\text{X}_1$:$(3, 11.5), \text{X}_2$:$(3, 10.75), \text{Y}_1$:$(5.25, 10.5), \text{Y}_2$:$(5.5, 10)$. }
\label{phase_diag_D2_2_14x14_NN_200av}
\end{center}
\end{figure*}
\subsection{Classification by Artificial Neural Network}
An artificial neural network (ANN) is a nonlinear function that takes an input ${\bf x}$ to compute an output ${\bf y}$ through sequential mappings by layers of ``neurons.''~\cite{nielsen_2015} A neuron itself is a nonlinear function that applies the activation function to each element of the weighted input ${\bf z} = W{\bf x}$, and a set of neurons that share the identical weight matrix is called a layer. In the following, we denote the operations corresponding to activation and weight matrix for the $i$th layer as $\mathcal{A}_i$ and $\mathcal{W}_i$,  respectively.  An ANN with layers that can be uniquely numbered according to the order of input and output and do not include any intralayer processing is referred to as a feedforward ANN. In this paper, we apply the feedforward ANN with two hidden layers. [See Fig. ~\ref{NN_architecture} for a graphical understanding of the architecture.]  The output is calculated as 
\begin{eqnarray}\label{NN}
{\bf y} = \mathcal{A}_3 \kl \mathcal{W}_3 \mathcal{A}_2 \kl \mathcal{W}_2 \mathcal{A}_1\kl \mathcal{W}_1{\bf x}\kr \kr \kr.
\end{eqnarray}
In our architecture, the activation function of the hidden and output layers are the Rectified Linear Unit (ReLU) and the Softmax function, respectively. The definitions are given as 
\begin{flalign}
{\rm ReLU}(z)&= {\rm max}(0,z) &\text{for } \mathcal{A}_1,\mathcal{A}_2,\\
{\rm Softmax}(z_j;{\bf z}) &= \exp(-z_j)/\sum_i \exp(-z_i)  &\text{for } \mathcal{A}_3.
\end{flalign}

Next, we discuss the training process of the machine. The parameters $\mathcal{W}$ are tuned via minimization of the cost function. This quantifies the performance of the machine which classifies the $\ztwo$ and the trivial phases in the binary classification scheme, and additionally the ThM phase in the ternary classification scheme. In a classification problem with the current network architecture, the cross-entropy function is widely used~\cite{bishop_2006} due to its convenience. In our paper, we also employ this function with the regularization term. Namely, the cost function is given as 
\begin{eqnarray}
\begin{split}
\Loss(\mathcal{W}) = &- \sum_{j=1}^{({\rm \#data})} \sum_{k=1}^{(\rm{\#class})} \hat{y}_j^{(k)} \log y_j^{(k)}({\bf x}_j;\mathcal{W})/({\rm \#data})\\
&+ \lambda \sum_{i =1}^{({\rm \# layers})} |\mathcal{W}^{(i)}|^2.
\end{split}
\end{eqnarray}
Here, $y^{(k)}_j$ is the output for the $k$-th label by the ANN, or ``the confidence of the machine'', for the $j$-th input training or test data, which is modified by the optimization of $\Loss$. On the other hand, $\hat{y}^{(k)} = \delta_{k.l_j}$ for the correct label $l_j$ denotes the corresponding phase for the data, and hence is constant throughout the training.
The second term, or the L2 regularization, suppresses the amplitude of the weight parameters, preventing the machine from overfitting to the training data.
The parameters are updated by mini-batch gradient descent with batch size 40 as \mbox{$\mathcal{W}^{(i)}_{j,k} \rightarrow \mathcal{W}_{j,k}^{(i)} -\eta \kl \partial \Loss/\partial \mathcal{W}_{j,k} \kr$}, where $\eta$ is the learning rate that is controlled by AdaGrad method to efficiently reach the global minimum.~\cite{duchi_2011} Furthermore, we apply the drop-out method to avoid overfitting.~\cite{srivastava_2014}

\subsection{Input data for machine}
Adopted as  the input data ${\bf x}$ is the disorder average over $N_r$ realizations of the spacial distribution of the quasiparticle, $P({\bf r})$, corresponding to the first excited state. Our expectation is that the qualitatively different behavior of the quasiparticle gives the machine sufficient information to discriminate phases. The bulk-edge correspondence in the $\ztwo$ phase, for instance, assures the robust edge-localization of the low-lying states across the zero energy. Furthermore, the behaviours in other two phases, namely the bulk-localization in the trivial phase and the delocalization of the quasiparticle over the system due to the bulk gap closing in the ThM phase, encourage us to consider $P({\bf r})$ for the lowest excitation as an appropriate input for the machine.

Let us consider the eigenstate $\ket{\psi}$ satisfying $H\ket{\psi} = E_1 \ket{\psi}$ with the lowest $E_1>0$. 
The degeneracy, if exists, is lifted up to time-reversal symmetry, and the two states are identical in terms of the quasiparticle distribution, namely,
\begin{eqnarray}\label{QP_dist}
P({\bf r}) = |\psi^e_{\ua}({\bf r})|^2 + |\psi^e_{\da}({\bf r})|^2 + |\psi^h_{\ua}({\bf r})|^2 + |\psi^h_{\da}({\bf r})|^2,
\end{eqnarray}
where the super(sub)script denotes the degree of freedom in the Nambu (spin) space. Some examples of   single disorder realization $P({\bf r})$ and its disorder average  $\braket{P({\bf r})}$ for $N_r = 500$ are shown in Fig. ~\ref{dist_av_ex}. 

While it is difficult to find evident pattern in respective $P({\bf r})$ due to the randomness, we expect that the translational symmetry is {\it statistically  recovered} by taking the disorder average. 
For instance, the bulk-edge correspondence assures the Majorana edge mode in the $\ztwo$ phase, which is robust against perturbation unless the bulk gap closes. 
The quasiparticle is localized at the edge although the amplitude of $P({\bf r})$ may become uneven along the circumference of the cylinder under spacial inhomogeneity. 
Such a fluctuation is eliminated by considering $\braket{P({\bf r})}$, which we confirm from the top row of Fig.~\ref{dist_av_ex}.
Furthermore, the localization in the bulk for the middle row indicates the thermal insulating property of the trivial phase, and the extension of the quasiparticle over the whole system in the bottom row reflects the metallic behavior of the ThM.

We classify the phases by feeding $\braket{P({\bf r})}$ to the ANN which learned the labels of $P({\bf r})$ in the clean limit. 
Both binary and ternary classification are considered.

\begin{figure*}[t]
\begin{center}
\includegraphics[width=1.95\columnwidth]{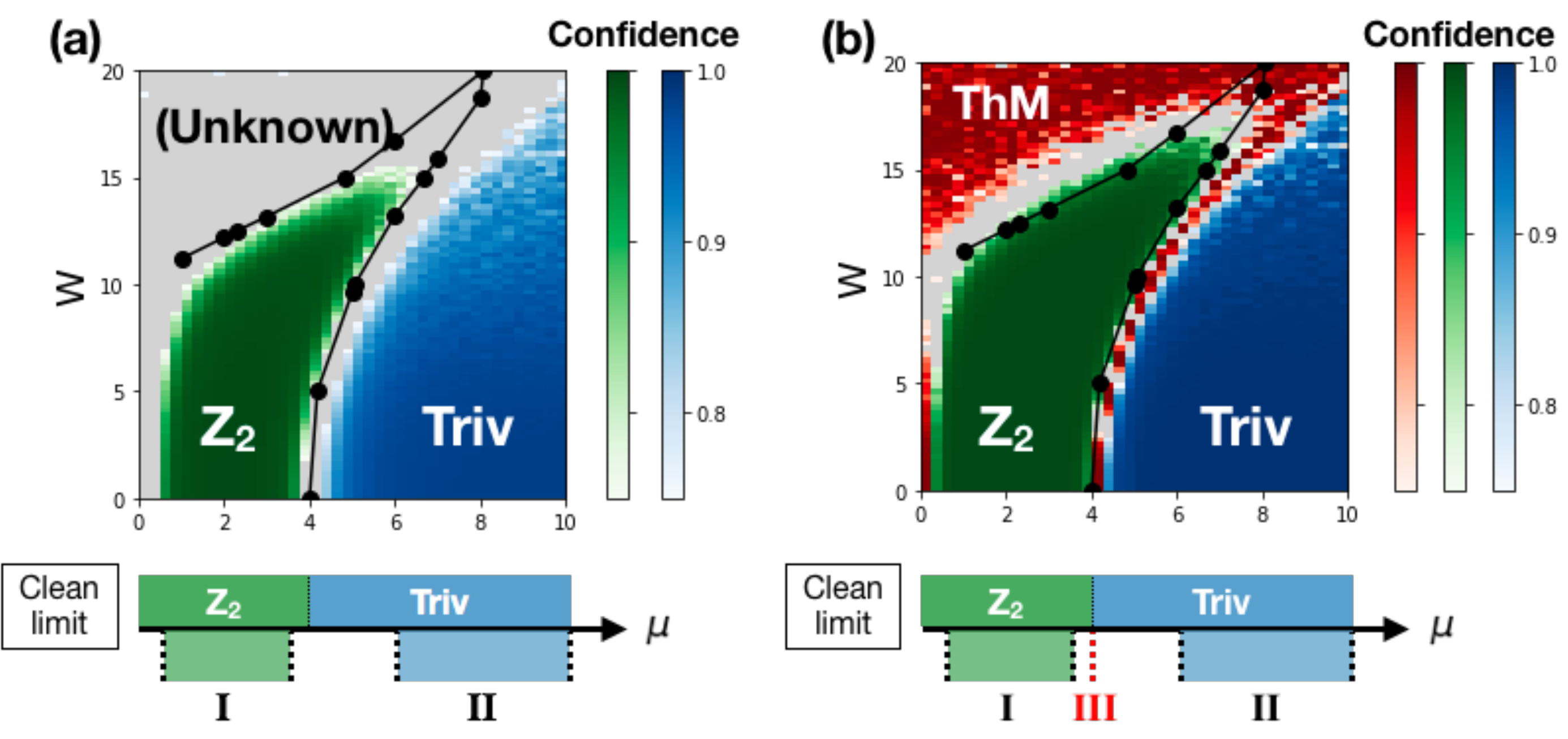}
\caption{(Color online) Average output of 200 binary-classifying ANNs trained with the clean-limit data for $t=1, \Delta = 3, \Delta_2 = 0$. 
The parameter $\mu$ of 1000 training data  with system size $14\times 14$ is uniformly distributed within (a) I: $[0.5,3.5]$ and II: $[6.0,10.0]$, (b) I, II, and III: $\mu = 0.0, 4.0$, each corresponding to the $\ztwo$ (green), the trivial (blue) phase and the critical point (red). 
The performance of the machine is monitored with the test data generated at $\mu \in [0.0,10.0]$ in the clean limit, resulting in over 95\% accuracy. The  outputs above 0.75 for $\braket{P({\bf r})}$ with $N_r$ = 500 are indicated by the depth of the color, and merely gray for below 0.75. }  
\label{phase_NN_clean_10x10_2phase_200av}
\end{center}
\end{figure*}
\section{Results}\label{results}
\subsection{Ternary classification}
First, we carry out the ternary classification at finite $\Delta_2$.
For $|\Delta_2| < |\Delta|,$ the bulk gap is closed when (i) $|\mu|< 2-2\sqrt{1-(\Delta_2/\Delta)^2}$ or (ii) $2+ 2\sqrt{1-(\Delta_2/\Delta)^2} < |\mu| < 4 \sqrt{1-(\Delta_2/\Delta)^2/2}$, and the system shows metallic behavior.~\cite{sato_2009,diez_2014} We focus on $\Delta = 3, \Delta_2 = 2$ and feed three phases, namely the $\ztwo$, trivial and ThM, to the machine,  expecting to predict the whole phase with high confidence. 
Shown in Fig.~\ref{phase_diag_D2_2_14x14_NN_200av} (a) is the average output of 200 ANNs which takes $\braket{P({\bf r})}$ with $N_r=$500 as the input. Each ANN is trained independently in a stochastic manner using the data from the clean limit indicated in Fig.~\ref{phase_diag_D2_2_14x14_NN_200av}(b). 
Only the region $\mu \geq0$ is shown since the phase diagram is symmetric with respect to $\mu = 0$. 
The black dots are the transition point obtained from the reliable TM method. [See Appendices A and B for details of other two methods.]
Remarkably the machine has successfully learned their characteristics even in the vicinity of the phase boundaries and fully extended the phase diagram. 
As is obvious from Figs.~\ref{phase_diag_D2_2_14x14_NN_200av}(c)-(d),  classifying  X$_1$ and X$_2$, or Y$_1$ and Y$_2$, with a comparable precision  is beyond our cognitive ability.

Next,  let us focus along $\mu = 2$. In the clean limit, the system is in the $\ztwo$ and enters the ThM and trivial phase sequentially by increasing the disorder, which is accurately captured by the ANN. The blurred output at $W \sim 15$ between the ThM and trivial phases is attributed to the larger fluctuation of the data, which is suppressed by increasing $N_r$. 
Other  $\ztwo$-ThM and ThM-trivial phase boundaries are nicely reproduced. 

Furthermore, the weak disorder region between the $\ztwo$ and the trivial phase at $\mu \sim 3.5$  is unambiguously classified as the ThM.
Let us emphasize
  again that this is attributed to the statistical recovery of the
  translational symmetry in the input data. Merely taking the average of the
  output is insufficient. [See Appendix C for further discussion.]
Note that such close parallel boundaries require extra effort on the other two methods; determining the  peak of the localization length, which diverges with the system size, by the TM becomes difficult due to the broadening by the finite-size effect, and that the noncommutative geometry approach does not work for critical phases.

\subsection{Binary classification}
To examine the binary classification by the ANN, we consider $\Delta_2 = 0$ at which the ThM phase is absent in the clean limit. 
Quasiparticle distributions are generated at $\mu \in [0.5,3.5]$ and $[6.0,10.0]$ for the $\ztwo$ and trivial phase, respectively.
The result is shown in Fig. ~\ref{phase_NN_clean_10x10_2phase_200av}.  As is expected, the machine reproduces the $\ztwo$-trivial phase boundary not only in the clean limit, i.e., the transition point $\mu = 4$, but also at $W>0$ which is obtained by the TM and  the noncommutative geometry approach.~\cite{MLTSC_comment2}

The drop of confidence along $\mu=0$ is also observed. This is understood as $\ztwo-\ztwo$ transition line, which is consistent with the analysis of the staggered fermion model  for class D.~\cite{medvedyeva_2010} Note that, such a transition that lacks the change in the size dependence on the thermal conductivity or localization length is very difficult to detect even by the TM method. 

The most remarkable confusion appears above the $\ztwo$ phase, e.g. $\mu=5$, which clearly suggests phase transition. [See the gray region in Fig. ~\ref{phase_NN_clean_10x10_2phase_200av}(a).]
While the output in the trivial phase at small disorder is close to unity, we observe that the confidence in the gray region is far below 1 regardless of the number of average for input or the machine.
 Such a confusion implies the qualitatively different feature  from the trivial phase, namely, the consequence of entering a completely different phase. 
To reinforce this argument,  we add two critical points , i.e., $\mu = 0$ for the $\ztwo-\ztwo$ and $\mu = 4$ for the $\ztwo-$trivial transition points, as the third label. We observe in Fig.~\ref{phase_NN_clean_10x10_2phase_200av}(b) that $\ztwo$-$\ztwo$ and $\ztwo$-trivial critical lines are present for finite $W$, and also that the previously confused region above the $\ztwo$  phase exhibits the extended behavior of the quasiparticle by the ANN. Hence,  this region is concluded as the thermal metal phase, which is also confirmed from the TM.  We note that the $\ztwo$-ThM phase boundary predicted by the machine is quantitatively consistent  with  the result by the numerical calculation of TM.

\section{Discussions}\label{discussions}
In this work, the use of the ANN to the classify phases of  2d noncentrosymmetric superconductor in class DIII with the disorder is shown to be valid in the following two cases. One is the extension of the phase diagram of $W=0$ to $W>0$ when all possible phases are present in the clean limit. We have confirmed that the machine successfully learns the property of each phase from the {\it quasi}-translational symmetric $\braket{P({\bf r})}$.  The confidence of the machine is high within the phases, which reflects the successful feature extraction.  Another is the detection of the unlearned phase. A correctly optimized ANN judges a state with high confidence when the learned feature is present in the data, and vice versa. The new phase does not exhibit localization in either bulk or the edge, and thus the machine is confused. 
We confirmed that in both cases the consistency with other independent methods holds.

Let us note that although the analysis here is based on the first moment of the quasiparticle distribution, in general, higher moments may also  play a crucial role.  In such a case, we expect that by adding the appropriate higher moments the classification can be done in other random system as well. 
Furthermore, we may consider alternative input to quasiparticle distribution for interacting systems with disorder;
as long as the quantity contains information on the system  and recovers the symmetry statistically, the validity of the proposed method is expected.
For instance,
  learning the entanglement spectra with the ANN has been shown to be a valid
  idea.\cite{evert_2017, shindler_2017, venderley_2017, evert_2017_2} Regarding the system without disorder, on the other hand,
  the auxiliary field configuration\cite{chng_2017, broecker_sign_2017} and the equal-time two-point
  correlation function,\cite{frank_2017} both obtained by Quantum Monte Carlo simulation,
  can be fed to the ANN to classify phases. We expect that such quantities are
  capable of capturing the property even when the disorder is present.

\section*{Acknowledgements}
The authors acknowledge helpful discussions with Hideaki Obuse, Masatoshi Sato, Ryo Tamura, and Shu Tanaka. This work was supported by JSPS KAKENHI Grant Nos. JP15K17719, JP16H00985, JP17K14352. N. Y. was supported by the  JSPS through Program for Leading Graduate Schools (ALPS) and JSPS fellowship (JSPS KAKENHI Grant No. JP17J00743).

\section*{Appendix A: Transfer Matrix}
\begin{figure*}[t]
\begin{center}

\begin{center}
\begin{tabular}{c}
  \begin{minipage}{0.33\hsize}
    \begin{center}
     \resizebox{!}{0.7\hsize}{\includegraphics{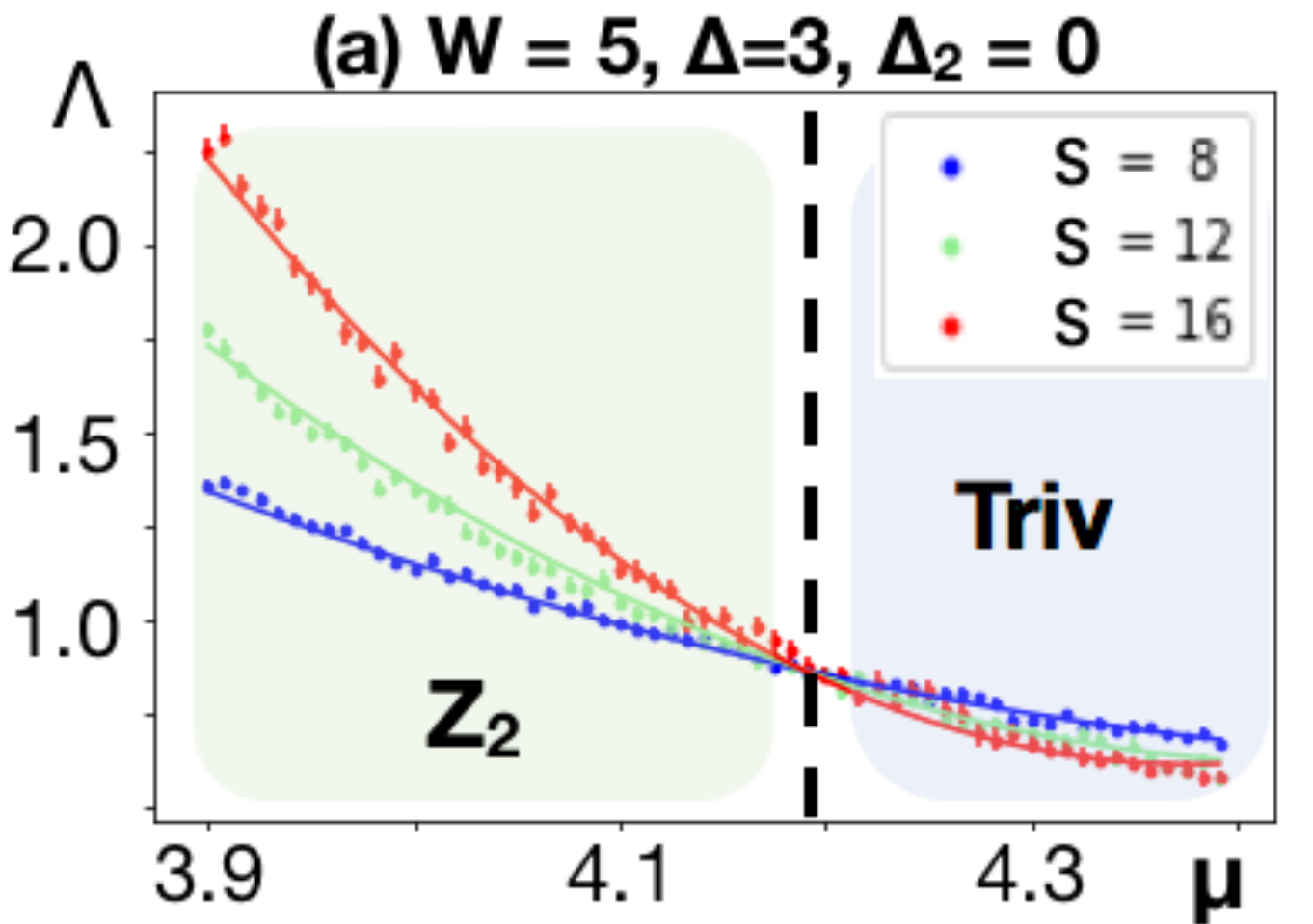}}
   \end{center}
  \end{minipage}
  \begin{minipage}{0.33\hsize}
   \begin{center}
     \resizebox{!}{0.7\hsize}{\includegraphics{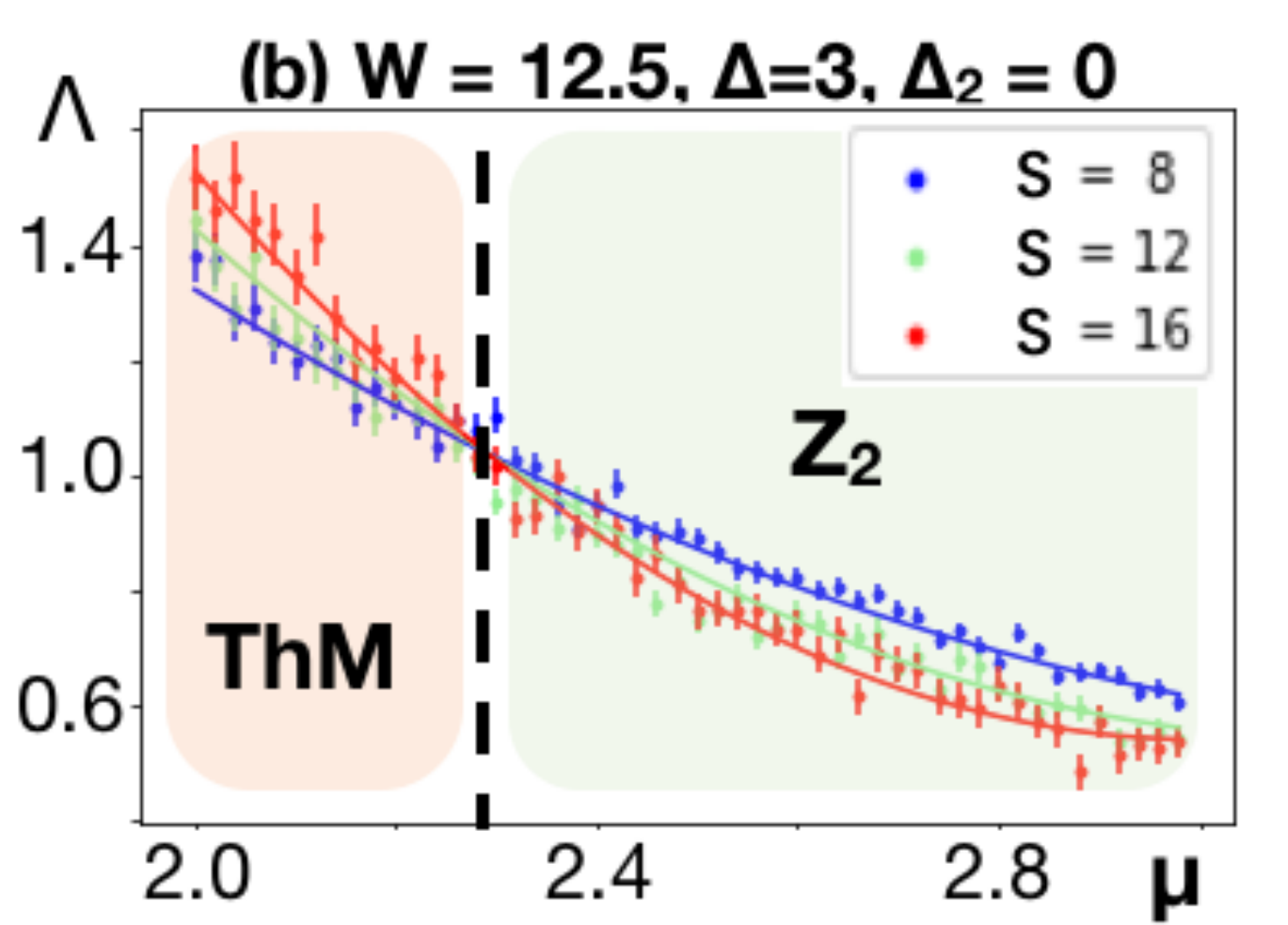}}
    \end{center}
  \end{minipage}
  \begin{minipage}{0.33\hsize}
   \begin{center}
     \resizebox{!}{0.7\hsize}{\includegraphics{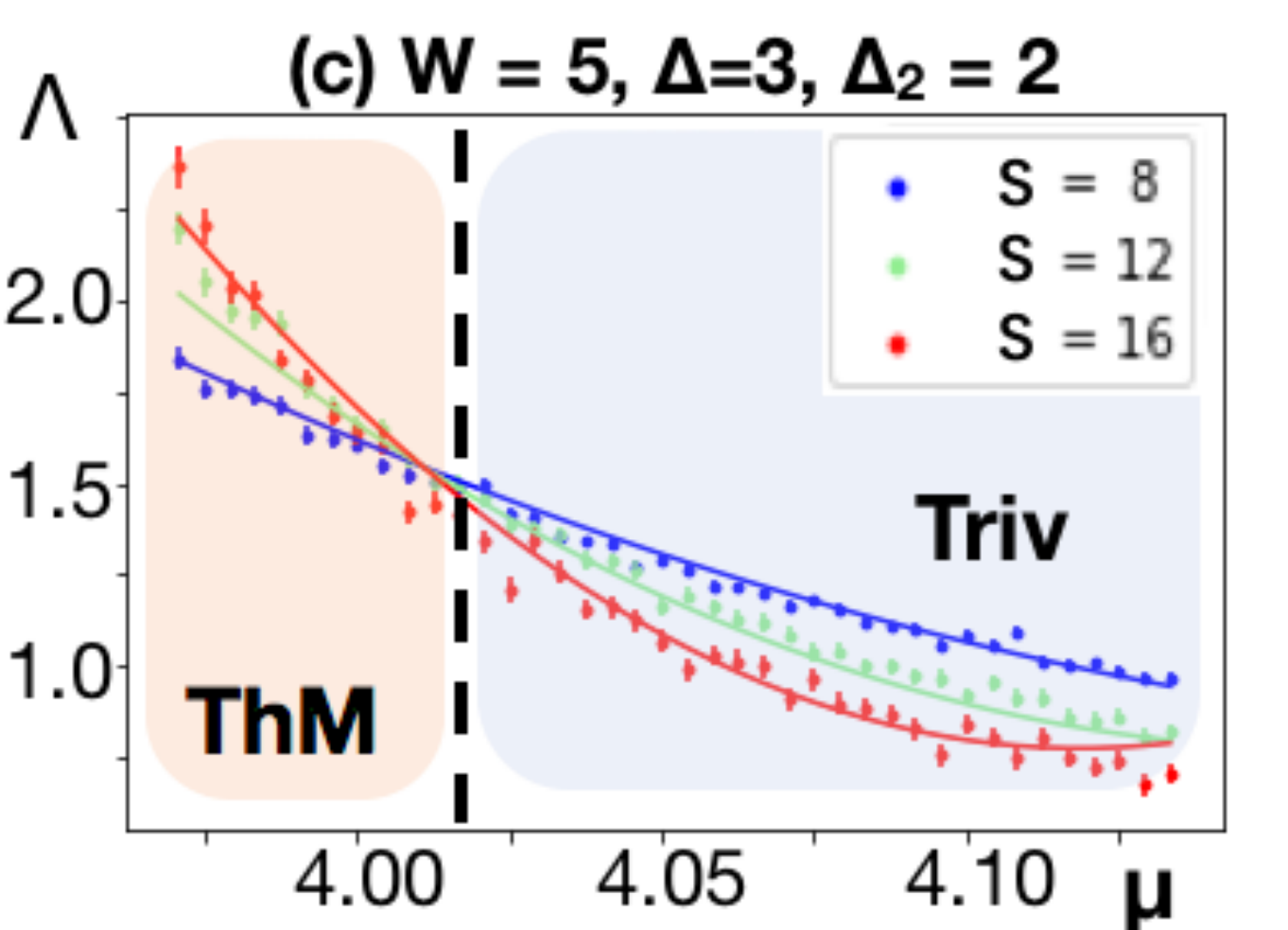}}
    \end{center}
  \end{minipage}
 \\
    \begin{minipage}{0.33\hsize}
    \begin{center}
     \resizebox{!}{0.8\hsize}{\includegraphics{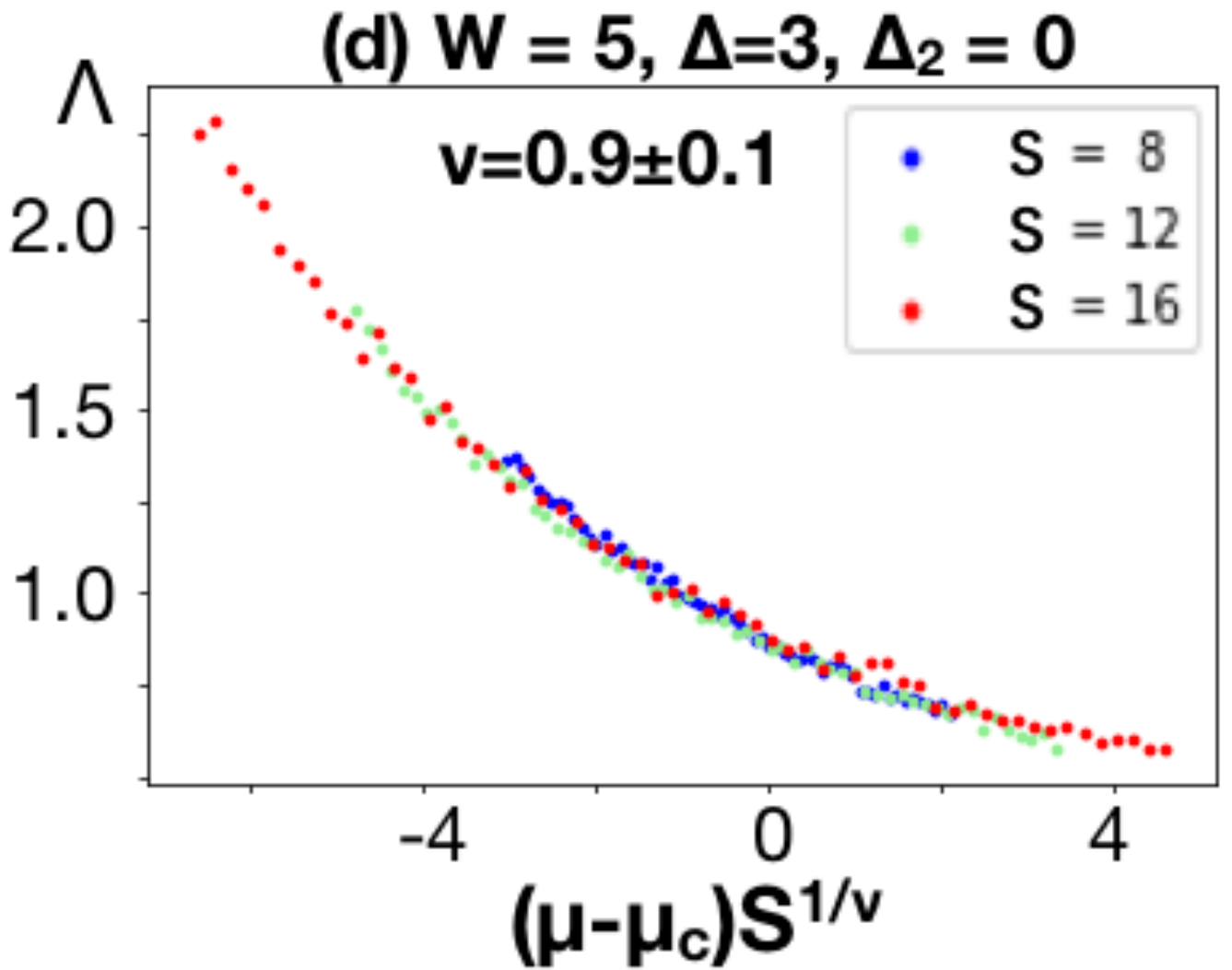}}
   \end{center}
  \end{minipage}
  \begin{minipage}{0.33\hsize}
   \begin{center}
     \resizebox{!}{0.8\hsize}{\includegraphics{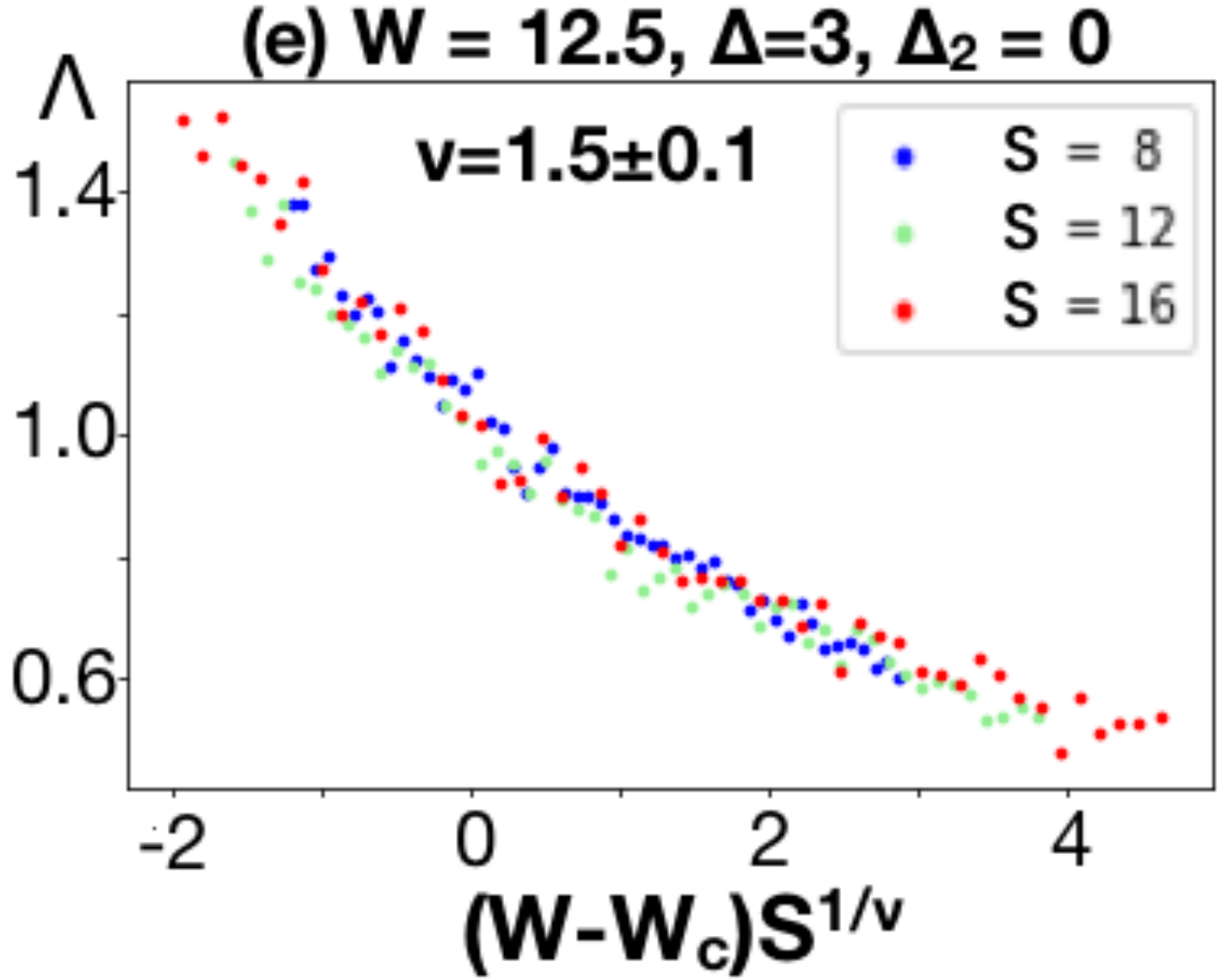}}
    \end{center}
  \end{minipage}
  \begin{minipage}{0.33\hsize}
   \begin{center}
     \resizebox{!}{0.8\hsize}{\includegraphics{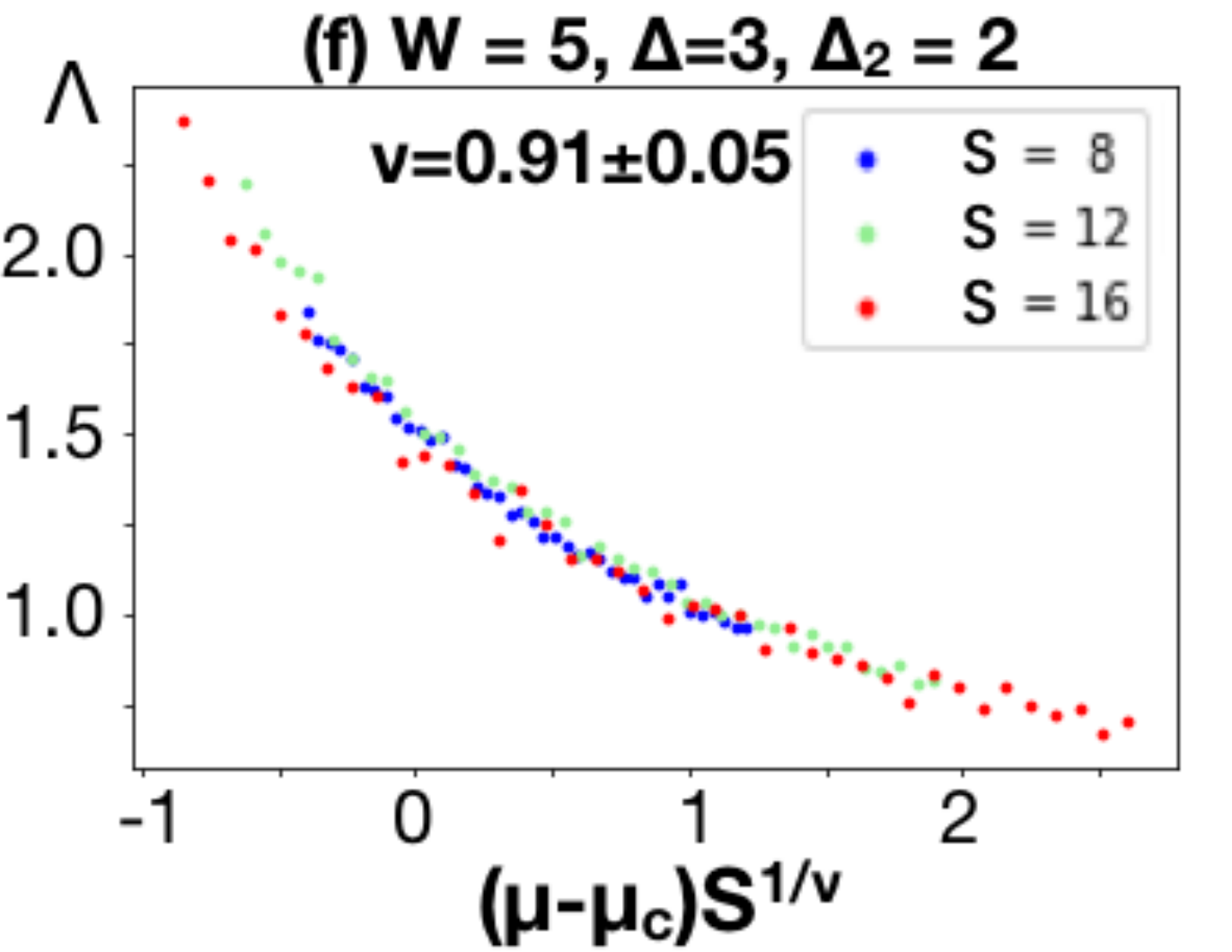}}
    \end{center}
  \end{minipage}

\end{tabular}
\end{center}

\caption{(Color online) Finite-size scaling of the dimensionless localization length, $\Lambda$, by the transfer matrix method for the Hamiltonian defined by Eqs. (1) - (5) in the main text is shown in (a)-(c). The parameters are given as $(W, \Delta, \Delta_2) = $ (a) $(5,3,0)$ under OBC, (b)  $(12.5, 3,0)$ under PBC, and (c) $(5,3,2)$ under OBC. Presented in (d)-(f) are the corresponding data collapses. The system width varies as $S = 8, 12, 16$. }  
\label{TM_example}
\end{center}
\end{figure*}
In this Appendix, we introduce the transfer matrix (TM) method for quasi-one-dimensional disordered system.\cite{mackinnon_1983, molinari_1997, yamakage_2013} 
Metal-insulator transition such as the Anderson transition can be understood from the size dependence of the localization length $\lambda$ and (thermal) conductivity $g$, which is easily computed by the TM method.
Let us consider a quasi-one-dimensional system with the length $L_x$ and the width $S$. We assume that the time-independent Schr\"odinger equation is  given as follows, 
\begin{eqnarray}\label{transfer_mat_ham}
L^{\dagger}_{i-1}{\boldsymbol \psi_{i-1}} + H_i {\boldsymbol \psi_i} + L_i{\boldsymbol \psi_{i+1}} = E {\boldsymbol \psi_i}.
\end{eqnarray}
Here, $H_i = H^{\dagger}_i$ is the Hamiltonian restricted on the $i$-th block and $E$ is the eigenenergy. We may simply consider the slice of the rectangle as a block when only the nearest neighbor hopping is present, but otherwise not necessarily a geometrical intersection. 
${\boldsymbol \psi_i}$ is the $2S$-dimensional wave function of the (quasi)particle on the $i$-th block and $L_i$ is the hopping matrix from $i$-th to $(i+1)$-th block. 
Assuming ${\rm det} |L_i| \neq 0$, Eq. ~(\ref{transfer_mat_ham}) is rewritten as
\begin{eqnarray}
\left(
\begin{array}{c}
{\boldsymbol \psi}_{i+1}\\
{\boldsymbol \psi}_i
\end{array}
\right) &=& 
\left(
\begin{array}{cc}
L_i^{-1}(E-H_i) & -L_i\\
I_{2S \times 2S} & 0
\end{array}
\right)
\left(
\begin{array}{c}
{\boldsymbol \psi}_i\\
{\boldsymbol \psi}_{i-1}
\end{array}
\right) \nonumber\\
&:=& \hat{T}_i(E)
\left(
\begin{array}{c}
{\boldsymbol \psi}_i\\
{\boldsymbol \psi}_{i-1}
\end{array}
\right)\label{tm_def},
\end{eqnarray}
where the above defined $\hat{T}_i(E)$ is referred to as one-step TM. The wave function at the edge and the total TM, $\hat{T}(E)$, is obtained as follows, 
\begin{eqnarray}
\left(
\begin{array}{c}
{\boldsymbol \psi}_{L_x}\\
{\boldsymbol \psi}_{L_x -1}
\end{array}
\right) =
\left( \prod_{i=0}^{L_x-1} \hat{T}_i \right)
\left(
\begin{array}{c}
{\boldsymbol \psi}_1\\
{\boldsymbol \psi}_0
\end{array}
\right) 
:= \hat{T}(E) \left(
\begin{array}{c}
{\boldsymbol \psi}_1\\
{\boldsymbol \psi}_0
\end{array}
\right).\label{tm_tot}
\end{eqnarray}
In the limit of $L_x\rightarrow \infty$, we consider the positive definite operator, $\displaystyle \hat{\Gamma} = \lim_{S\rightarrow \infty}(\hat{T}\hat{T}^{\dagger})^{1/2S}$, to introduce
\begin{eqnarray}
\lambda_j = \frac{1}{\ln \gamma_j},
\end{eqnarray}
where $\gamma_j$ is the $j$-th eigenvalue, which is positive and finite, of $\hat{\Gamma}$.
The corresponding eigenfunction behaves as $\exp(\pm x/\lambda_j)$, with the sign denoting the direction of the decay, and therefore $\lambda_j$ can be  understood as the localization length. We set the length of the system from $10^4$ to $10^5$ so that the statistical error is small enough.

As pointed out by MacKinnon and Kramer, the finite-size scaling of the maximum localization length, $\lambda_{\rm max} := \lambda$, is equivalent to the scaling theory of conductance $g$.\cite{abrahams_1979, mackinnon_1983} 
The dimensionless localization length in the vicinity of the metal-insulator transition point is assumed to be expressed by one-parameter scaling. Namely, by writing the parameter related to the transition as $q$ (e.g. chemical potential $\mu$ and the amplitude of the Anderson potential $W$ in our work), 
\begin{eqnarray}
\Lambda(q) := \frac{\lambda(q)}{S}
&=& \Lambda_c + \sum_{n=1}^{N}a_n (q-q_c)^n S^{n/\nu} \nonumber\\
&+& \sum_{n=1}^{N'}b_n(q-q_c)^n S^{n/\nu + y},
\end{eqnarray}
where the subscripts $c$ denote the value at the critical point, $a_n$ and $b_n$ are the expansion coefficients, and $\nu$ is the critical exponent for the localization length. The third term is the irrelevant length scale collection by the boundary, whose size dependence corrected by $y<0$. Finite integers $N$ and $N'$ denote the number of the fitting parameters, which is taken as $N=2, N'=0$ in this work. Examples for $\ztwo$-trivial and $\ztwo$-thermal metal (ThM) phase transitions are shown in Fig. \ref{TM_example}, in which the rising(falling) of $\Lambda$ for extended(localized) states are indeed observed.

Last but not least, let us note that appropriate boundary condition must be applied to detect the transition from or into the $\ztwo$ phase.\cite{yamakage_2013} In two-dimensional systems, we have two options: open and periodic boundary condition (OBC or PBC) along the direction perpendicular to the transferred direction. 
The edge state appears along the transferred direction with the OBC, while the state is merely localized in the first or the last block with the OBC. Thus, to determine $\ztwo$-trivial ($\ztwo$-ThM) phase boundary, we must consider OBC (PBC) system. Note that the trivial-ThM boundary is detected in either way.

\section*{Appendix B: ${\mathbb Z}_2$ index for 2d class DIII system with disorder}
\begin{figure*}[ht]
\begin{center}

     \resizebox{0.9\hsize}{!}{\includegraphics{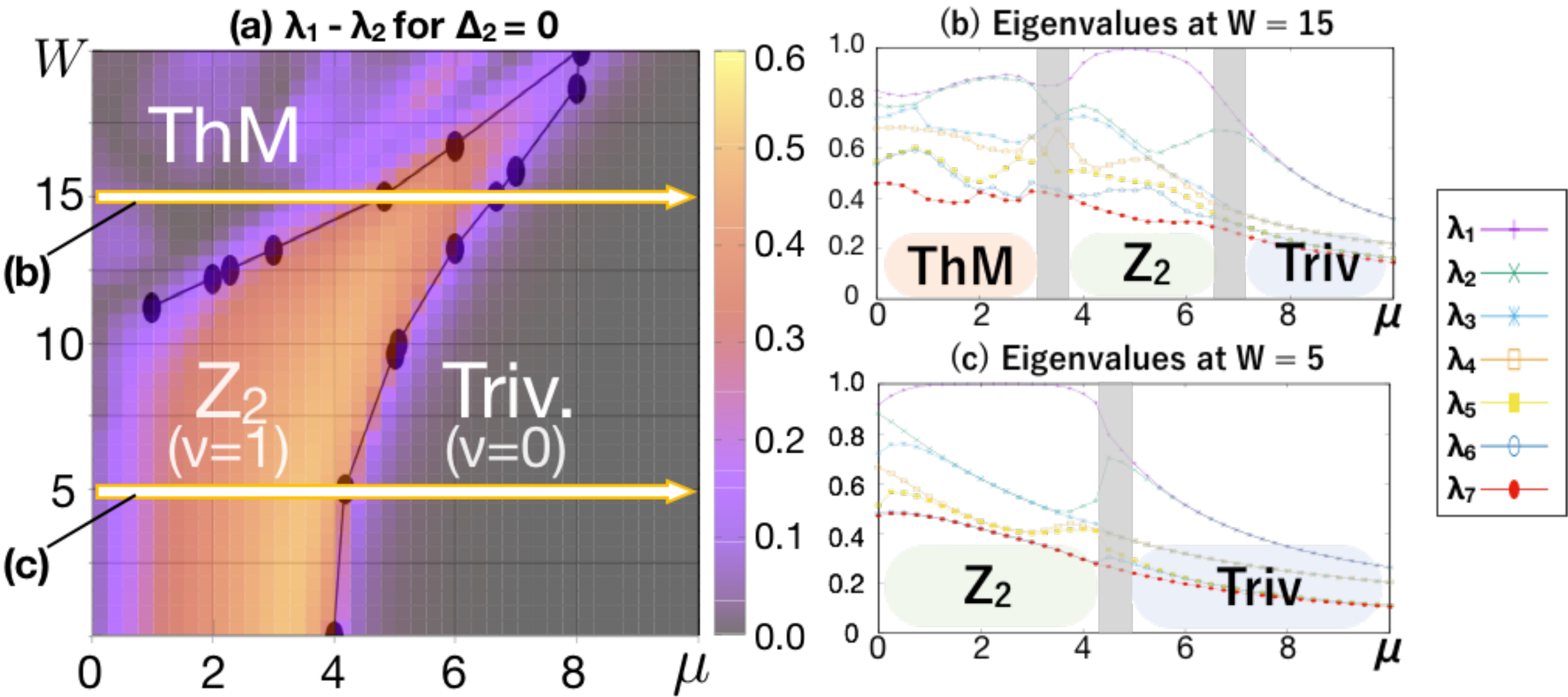}}
     \resizebox{0.9\hsize}{!}{\includegraphics{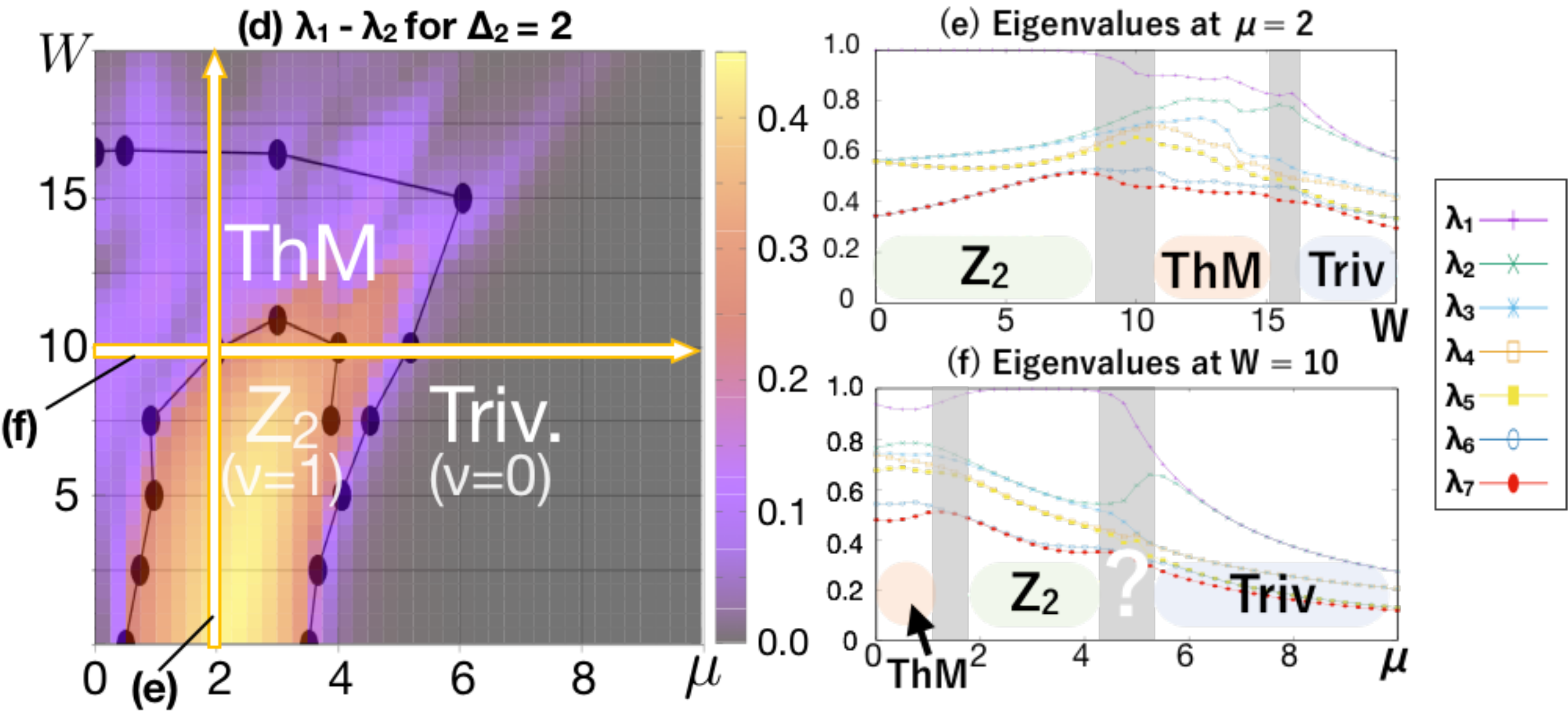}}

\caption{(Color online) (a) and (d) $\lambda_1- \lambda_2$ as a function of the chemical potential $\mu$ and the disorder strength $W$ for class DIII Hamiltonian given by Eq.~(1)-(5) of the main text.
The parameters are taken as $(t, \Delta, \Delta_2) = $ (a) \mbox{(1, 3, 0)} and (d) \mbox{(1, 3, 2)}, respectively, and the system size is \mbox{$20 \times 20$}.
(b), (c), (e), and (f) show $\mu$ and $W$ dependences of the eigenvalues $\lambda_i$, $i=1,2,\ldots,7$, of the operator ${\cal A}$ for the system size \mbox{$20 \times 20$}.
The parameters are taken as $(t, \Delta, \Delta_2, W) = $ (b) \mbox{(1, 3, 0, 15)}, (c) \mbox{(1, 3, 0, 5)}, (f) \mbox{(1, 3, 2, 10)}, 
and (e) $(t, \Delta, \Delta_2, \mu) = $ \mbox{(1, 3, 2, 2)}, respectively.
The gray bars in (b), (c), (e), and (f) denote marginal areas.
}  
\label{NCI_phase_diag}
\end{center}
\end{figure*}

In this appendix, we introduce the noncommutative geometry approach to map out the phase diagram of 2d class DIII system. The $\ztwo$ index 
derived in previous works~\cite{katsura_2016_1, katsura_2016_2} is numerically advantageous since it can be determined from the discrete spectrum of a certain compact operator without taking the disorder average. 
See Ref. [43] for detailed 
numerical implementation. 
The definition of the $\ztwo$ index of 2d class DIII system is given as 
\begin{eqnarray}
\nu = {\rm ker\  dim} \ [\mathcal{A}-1]\ \ {\rm modulo} \ 2,
\end{eqnarray}
where $\nu = 0$ and $1$ correspond to the trivial and the $\ztwo$ phases, respectively.  The operator $\mathcal{A}$ measures the difference 
between two projections,
\begin{eqnarray}
{\cal A} = P_{\rm F} - \mathcal{D}_a^* P_{\rm F} \mathcal{D}_a.
\end{eqnarray}
Here, $P_{\rm F}$ is the projection operator onto the quasiparticle states below zero energy. 
The Dirac operator $\mathcal{D}_a$ is defined by
\begin{eqnarray}
\mathcal{D}_a({\bm r}) :=  \frac{r_1 + i r_2 - (a_1 + i a_2)}{|r_1 + i r_2 - (a_1 + i a_2)|},
\end{eqnarray}
where ${\bm r} = (r_1, r_2) \in \mathbb{Z}^2$ denotes the position operator of a square lattice and ${\bm a} = (a_1, a_2) \in \mathbb{R}^2\backslash \mathbb{Z}^2$ is a vector off the lattice points.
The operator $\mathcal{D}_a^*$ is the adjoint of the Dirac operator $\mathcal{D}_a$.
Hereafter, we regard $\lambda_i$ as the {\it i}-th eigenvalue of the operator $\mathcal{A}$ 
in descending order including multiplicity.

\begin{figure*}[t]
\begin{center}

\begin{center}
\begin{tabular}{c}
  \begin{minipage}{0.42\hsize}
    \begin{center} \hspace{-0.5cm}

     \resizebox{!}{0.95\hsize}{\includegraphics{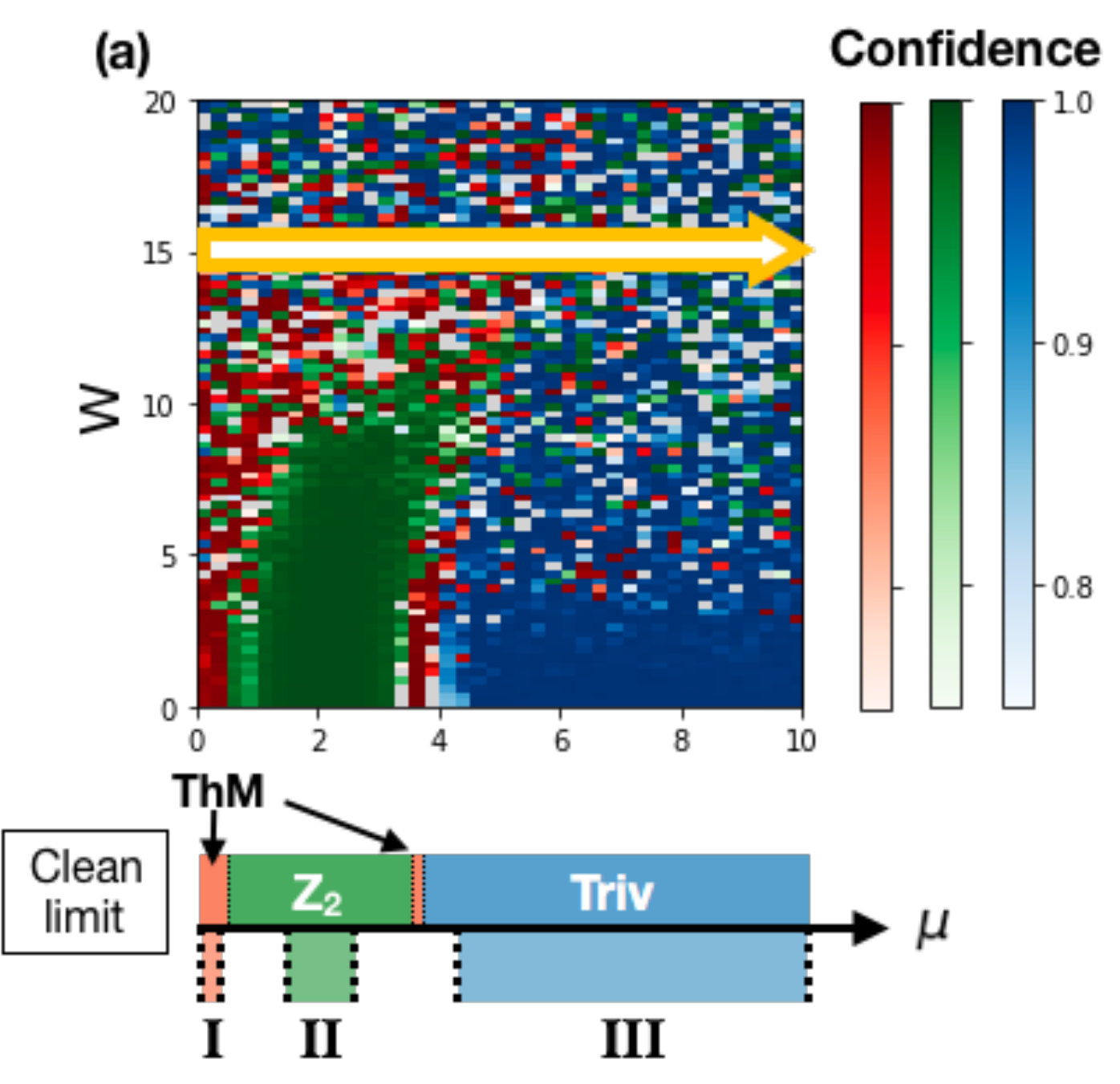}}
   \end{center}
  \end{minipage}\hspace{1.4cm}
  \begin{minipage}{0.52\hsize}
   \begin{center}
     \resizebox{!}{0.99\hsize}{\includegraphics{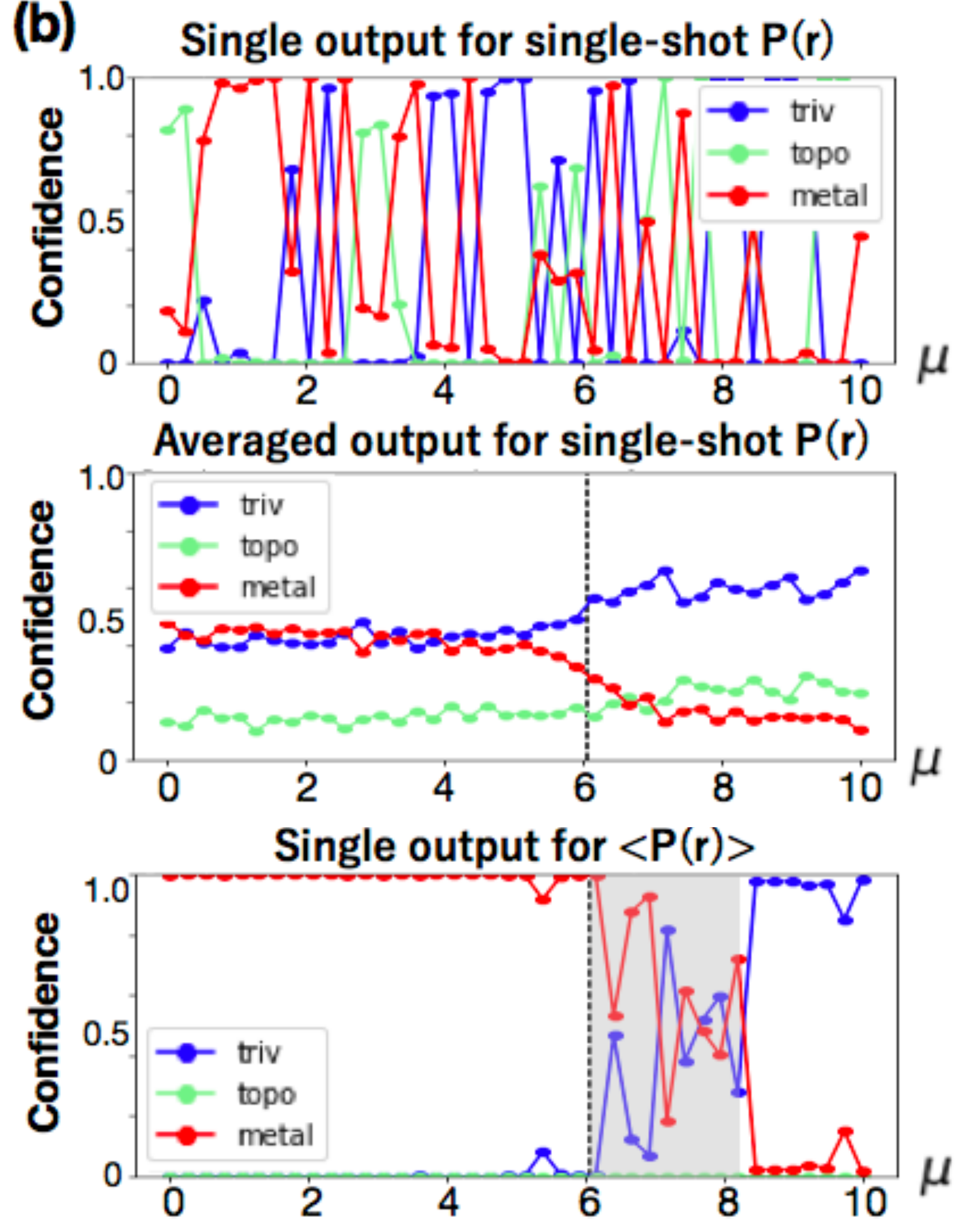}}
    \end{center}
  \end{minipage}

\end{tabular}
\end{center}

\caption{(Color online) (a) The output of ANN for single-shot $P(\bf r)$ for $\Delta=3, \Delta_2 = 2$. Boundary between phases are hardly recognizable. (b) The single output for $P({\bf r})$, the average of 200 outputs for independently generated $P({\bf r})$, and the single output for $\braket{P(\bf r)}$ with $N_r = 500$ from the top. The amplitude of the random potential is fixed as $W=15$.}  
\label{ANN_singleshot}
\end{center}
\end{figure*}


Shown in Fig.~\ref{NCI_phase_diag}(a) is $\lambda_1-\lambda_2$ as a function of the chemical potential $\mu$ and the disorder amplitude $W$ with the pairings fixed as $\Delta=3$ and $\Delta_2 = 0$. The orange-colored region denotes the $\ztwo$ phase since  $\lambda_1 \sim 1$ [see, for instance, Fig.~\ref{NCI_phase_diag}(c)] and $\lambda_1 - \lambda_2 \neq 0$ evidently hold and thus imply $\nu =1$. In Fig.~\ref{NCI_phase_diag}(a) we see that the numerical result is in good agreement with the boundary obtained by the TM. The two black areas above and to the right of the $\ztwo$ phase are identified as the ThM and the trivial phases, respectively. This is done in the following way. When the spectral gap is open ($=$ trivial or $\ztwo$ phase), 
the eigenvalues below unity always come in pairs [see Fig.~\ref{NCI_phase_diag}(b)(c)] 
owing to the two symmetries: the time-reversal symmetry of the Hamiltonian and the supersymmetric structure of the operator $\mathcal{A}$.
However, 
the doublet structure is not guaranteed when the spectral or the mobility gap vanishes (ThM phase), and in fact, each eigenvalue shows no such a specific structure 
in the leftmost region of Fig.~\ref{NCI_phase_diag}(b). 


The difference 
between the first and 
second eigenvalues 
for $\Delta_2 = 2$ is also given in Fig.~\ref{NCI_phase_diag}(d).
In the orange region, $\lambda_1 \sim 1$ [see, for instance, Fig.~\ref{NCI_phase_diag}(e)] and $\lambda_1 - \lambda_2 \neq 0$, and hence $\nu = 1$ which corresponds to the $\ztwo$ phase.
The black region denotes the trivial phase with $\nu = 0$ because there is no $\lambda_1 \sim 1$. [See Fig.~\ref{NCI_phase_diag}(e)]. 
While the boundary of the $\ztwo$ phase is consistent with the TM, detection of the phase boundary between the ThM and the trivial phase requires deep consideration in some situations. 
In Fig.~\ref{NCI_phase_diag}(e), the two phases are distinguishable by the presence of the doublet structure, 
whereas in Fig.~\ref{NCI_phase_diag}(f), it is hard to tell whether the intermediate region between the $\ztwo$ and the trivial phase is a finite window of the ThM. 
As is seen in Fig.~3 of the main text, this is indeed a small window of ThM, which is unambiguously captured by the ANN.

\section*{Appendix C: Single-shot and averaged data}

In the following, we see that the success by the ANN is attributed to the recovery of symmetry, but not merely by the law of large numbers. Taking disorder average of the input data corresponds to an appropriate feature selection, which is crucial in training our machine. Since the ANN is a totally nonlinear function, this is not the case for averaging the output.

As is shown in Fig. \ref{ANN_singleshot}(a), classification of $P({\bf r})$, i.e., the single-shot realization, results in a total meaninglessness, particularly in the strong disorder region. For the sake of simplicity,  let us restrict the amplitude as $W=15$ in the following. Shown at the top of Fig. \ref{ANN_singleshot}(b) is the output for single-shot. The random values reflect the fact that the ANN is confused by the translational-symmetry-broken behavior of the quasiparticle. We see in the middle that averaging such outputs in a brute-force manner does not improve the situation at all. Although the  faint slope around the boundary seems to capture the phase transition, the output converges far below the unity. It is questionable whether we can determine the phase in general. Shown at the bottom is the appropriate classification for $\braket{P({\bf r})}$ with $N_r = 500$, in which the feature of the quasi-translational states  are detected appropriately.

\bibliographystyle{apsrev4-1}
\bibliography{readcube_export}

\begin{thebibliography}{70}
\bibitem [{MLT({\natexlab{a}})}]{MLTSC_comment6}%
  \BibitemOpen
  \href@noop {} {\bibfield  {journal} {\bibinfo  {journal} {Another well-known
  issue is the increase in the numerical cost. Together with the necessity to avoid
  overfitting, this problem implies that excessive pursuit in decreasing
  $\epsilon$ is undesirable in practical situations}}
 }\BibitemShut {NoStop}%
\bibitem [{MLT({\natexlab{c}})}]{MLTSC_comment1}%
  \BibitemOpen
  \href@noop {} {\bibfield  {journal} {\bibinfo  {journal} {{Although the
  square lattice is considered in the effective model, i.e. Eq. (1), we assume
  that the underlying crystal structure of the original Hamiltonian breaks
  inversion symmetry}}} }\BibitemShut {NoStop}%
\bibitem [{MLT({\natexlab{d}})}]{MLTSC_comment2}%
  \BibitemOpen
  \href@noop {} {\bibfield  {journal} {\bibinfo  {journal} {{At larger
  disorder, the phase boundaries approach each other. Therefore, machine is
  confused, i.e, the output remains far below 1, by the finite-size effect,
  resulting in the small estimation of the $\ztwo$ phase}}}
}\BibitemShut {NoStop}%
\end{thebibliography}

\end{document}